\documentclass[12pt,epsfig]{article}

\usepackage{amssymb,amsmath}
\usepackage{epsfig,float}
\usepackage[hypertex,colorlinks=true,linkcolor=red,citecolor=blue]{hyperref}

\newcommand{\be}{\begin{eqnarray}}
\newcommand{\ee}{\end{eqnarray}}
\newcommand{\ba}{\begin{array}}
\newcommand{\ea}{\end{array}}

\newcommand{\bea}{\begin{eqnarray}}
\newcommand{\eea}{\end{eqnarray}}

\newcommand{\bi}{\begin{itemize}}
\newcommand{\ei}{\end{itemize}}

\begin{document}
\begin{titlepage}

\begin{flushright}
\begin{tabular}{l}
\end{tabular}
\end{flushright}
\vspace{1.5cm}

\begin{center}
{\LARGE \bf
Differential technique for the covariant orbital angular momentum
operators
}
\vspace{1cm}

\renewcommand{\thefootnote}{\alph{footnote}}

{\sc M.~Matveev}${}^{1}$,
{\sc A.~Sarantsev}${}^{1,2}$,
{\sc K.~Semenov-Tian-Shansky}${}^1$,
{\sc A.~Semenova}${}^1$,
%
\\[0.5cm]
\vspace*{0.1cm} ${}^1${\it
National Research Centre ``Kurchatov Institute'': Petersburg Nuclear Physics
Institute, RU-188300 Gatchina, Russia
                       } \\[0.2cm]
\vspace*{0.1cm} ${}^2$ {\it
Helmholtz-Institut f\"{u}r Strahlen- und Kernphysik der Universit\"{a}t Bonn, Nu{\ss}allee 14-16, D-53115 Bonn, Germany
                       } \\[0.2cm]
{\it \large \today
 }
\vskip2cm
{\bf Abstract:\\[10pt]} \parbox[t]{\textwidth}{

The orbital angular momentum operator expansion turns to be
a powerful tool to construct the fully covariant
partial wave amplitudes of hadron decay reactions and hadron
photo- and electroproduction
processes.

In this paper we consider a useful development of the
orbital angular momentum operator expansion method. We present
the differential technique allowing the
direct calculation of convolutions of two orbital angular momentum operators
with an arbitrary number of open Lorentz indices. This
differential technique greatly simplifies calculations when
the reaction subject to the partial wave analysis involves high spin
particles in the initial and/or final states. We also present a useful
generalization of the orbital angular momentum operators.
  }
\vskip1cm
\end{center}

\vspace*{1cm}
\end{titlepage}

\setcounter{footnote}{0}

\section{Introduction}
\mbox


Study of the strong interaction at low and intermediate energies
provides us the detailed information on the spectrum, properties
and structure of strongly interacting particles - hadrons.
This knowledge is indispensable for improving our understanding
of QCD in the nonperturbative regime. It also brings crucial
tests both for the QCD-inspired phenomenological strong interaction models and for
the lattice QCD calculations.

The resent discoveries in the heavy quark sector
\cite{Ablikim:2013mio,Ablikim:2013wzq,Aaij:2015eva,Aaij:2015tga}
brought evidences for the existence of the tetraquark mesons and pentaquark baryons
(see {\it e.g.} \cite{Karliner:2017qhf} for a recent review and
complete set of references). This definitely forces us to go beyond
the naive quark model that describes mesons and baryons as bound
states of quark-antiquark and three quarks  respectively. In
the light quark meson sector the situation is yet not so clear,
however a number of collaborations have reported the observation of
the so-called {\it exotic mesons} with quantum numbers forbidden for
the
$q \bar{q}$-system. For the light quark baryon sector it
worths mentioning the long standing problem
\cite{Dalitz:1959dn}
of structure of $\Lambda(1405)$ $J^P=\frac{1}{2}^-$ resonance which is
considered to be a candidate for the $\bar{K}N$ bound state
\cite{Jido:2003cb, Mai:2012dt}. Another example is the controversial
$uudd \bar{s}$ $\theta^+(1540)$ state \cite{Diakonov:1997mm} widely
discussed in the 2000s (see \cite{Danilov:2007bp} for a discussion)
and its non-strange partners (see {\it e.g.}
\cite{Kuznetsov:2017qmo} and references therein).



A usual pattern for the exotic states decays are
the channels involving three or more particles in the final state.
This also turns to be the case for the radial
excitations of the non-exotic states.
Signals from such
states can be extracted from experimental data
by means of a rather involved partial
wave analysis. This analysis must correctly
account for all correlations in the multidimensional phase
space between amplitudes corresponding to all possible decay chains.

In a number of analyzes 
(see {\it e.g.} \cite{Aaij:2015tga}) 
the corresponding partial wave amplitudes
are constructed
by means of the two step procedure. Firstly, one calculates amplitudes for 
transition between all possible two particle channels. The angular
dependence of such amplitudes is described by the well known
non-relativistic spherical functions. At the second step, the
complete amplitude for a given decay chain is constructed as a
product of the corresponding two-particle amplitudes, which are
rotated and boosted into a particular reference frame.  Then, the
amplitudes for all possible decay chains, in turn, are rotated and
boosted to one selected reference frame to get the correct
interference pattern. However, this approach is
plagued with ambiguities in the boosting procedure. It seems that a
more straightforward way is to use the covariant approach from the
very beginning. By construction, the covariant approach  is
independent of the reference frame and can be directly applied to
any reaction with multiparticle final states.

The development of the 
covariant approach for the partial wave analysis of the experimental
data has a long history
\cite{Zemach:1968zz,Scadron:1968zz,Alfaro_red_book,Chung:1971ri,Zou:2002ar}.
The systematic method, which allow  to
construct partial wave amplitudes for arbitrary value of
resonance spin and arbitrary spins of
decay particles, was suggested in
\cite{Anisovich:2001ra}
and saw further development in
\cite{Anisovich:2004zz,Anisovich:2006bc}.
It is based on the spin-orbital classification of  partial wave amplitudes.
One of the main components of this approach is the construction of the
orbital-angular-momentum-operators (OAM-operators)
$X^{(n)}_{\mu_1 \ldots \mu_n}$.
The  OAM-operators are the Lorentz tensors
with rank which corresponds to the orbital angular momentum
$n$ of the resonating two particle system.
These operators can be constructed with the help of recurrent relations expressing
higher order  OAM-operators through the operators of lower ranks.




Let us consider a resonance decay into an intermediate state and a spectator state
of particular orbital angular momentum.
The corresponding partial wave amplitudes are
constructed as convolutions of several OAM-operators.
These convolutions may have open Lorentz indices,
which are further convoluted with the polarization
vectors of the initial and final state particles.
%
%
Calculations of OAM-operators convolutions with open Lorentz
indices, in many cases, represent
a rather complicated mathematical task.

In this paper we present the differential technique allowing a
direct calculation of convolutions of two OAM-operators
with an arbitrary number of open Lorentz indices.
We consider this procedure as a significant step in the
development of the 
covariant approach for the
partial wave analysis of the reactions with multiparticle final
states. It also will help to develop the fully
electromagnetically gauge invariant description
of hadron photo- and electroproduction reactions.

\label{Sec_Intro}
\mbox

\section{Properties of orbital-angular-momentum-operators
$X^{(n)}_{\mu_1 \ldots \mu_n}$}
\label{Sec_OAM_prop}
\mbox

In this Section we review the basic properties of the the
OAM-operators
$X^{(n)}_{\mu_1 \ldots \mu_n}$.
These operators occur in the description of the
decay of a composite particle
with integer spin
$n$
and momentum
$P=k_1+k_2$ ($P^2=s$)
into two spinless particles with momenta
$k_1$
and
$k_2$.
They also serve as building blocks for the description
of more involved cases.

In order to ensure that the operators
$X^{(n)}_{\mu_1 \ldots \mu_n}$
correspond to the appropriate
spin-$n$ irreducible representations of the
Lorentz group these operators should satisfy the
following list of properties \cite{Anisovich:2001ra}:
\bi
\item Symmetry with respect to permutation of indices:
\be
X^{(n)}_{\mu_1 \ldots \mu_i \ldots \mu_j \ldots \mu_n}=
X^{(n)}_{\mu_1 \ldots \mu_j \ldots \mu_i \ldots \mu_n};
\label{Prop_sym}
\ee
\item Orthogonality to the total momentum $P$:
\be
P^{\mu_i}X^{(n)}_{\mu_1 \ldots \mu_i \ldots  \mu_n}=0;
\label{Prop_orth}
\ee
\item The tracelessness property over any pair of indices:
\be
g^{\mu_i \mu_j} X^{(n)}_{\mu_1 \ldots \mu_i \ldots \mu_j \ldots \mu_n}=0.
\label{Prop_trace}
\ee
\ei
In order to satisfy the orthogonality condition 
(\ref{Prop_orth})
the
OAM-operators are constructed from
the relative momentum
$k^\bot_\mu$
and the orthogonal metric tensor
$g^\bot_{\mu \nu}$:
\be
k^\bot_\mu= g^\bot_{\mu \nu}\, \frac{1}{2}(k_1-k_2)^\nu;
\ \ \
g^\bot_{\mu \nu}= g_{\mu \nu}-\frac{P_\mu P_\nu}{s}. 
\label{metric}
\ee
Note that the trace of the orthogonal metric tensor
$g^\bot_{\mu \nu}$ defined in
(\ref{metric})
is
$$
g^{\bot \; \mu}_{\mu} \equiv 3.
$$

The operator for $n=0$ is a scalar and the $n=1$
operator is just $k^\bot_\mu$:
\be
X^{(0)}(k_\bot)=1; \ \ \ X^{(1)}_{\mu}(k_\bot)=k^\bot_\mu.
\ee
The operators $X^{(n)}_{\mu_1 \ldots \mu_n}(k_\bot)$
for $n >1$ can be constructed from the recurrence relation
\be
&&
X^{(n)}_{\mu_1 \ldots \mu_{n}}(k_\bot)=k_\bot^\alpha Z^{(n-1)}_{\mu_1 \ldots \mu_n, \alpha},
\ \ \ {\rm where}
\nonumber \\ &&
Z^{(n-1)}_{\mu_1 \ldots \mu_n, \alpha}=
\frac{2n-1}{n^2} \sum_{i=1}^{n} g^\bot_{\mu_i \alpha}
X^{(n-1)}_{\mu_1 \ldots \mu_{i-1}  \mu_{i+1} \ldots \mu_{n} }(k_\bot)
\nonumber \\ &&
-\frac{2}{n^2}
\sum_{i,j=1 \atop i<j}^{n} g^\bot_{\mu_i \mu_j}
X^{(n-1)}_{\mu_1 \ldots \mu_{i-1}  \mu_{i+1} \ldots  \mu_{j-1}  \mu_{j+1} \ldots\mu_{n} \alpha}(k_\bot).
\label{Recurrence_rel}
\ee

It is straightforward
to check that
$X^{(n)}_{\mu_1 \ldots \mu_{n}}(k_\bot)$
defined from
(\ref{Recurrence_rel}) satisfies the properties listed
in
Eqs.~(\ref{Prop_sym}), (\ref{Prop_orth}), (\ref{Prop_trace}).

The following convolution identity is valid
\be
k_\bot^{\mu_n} X^{(n)}_{\mu_1 \ldots \mu_{n}}(k_\bot)=
|k_\bot|^2 X^{(n-1)}_{\mu_1 \ldots \mu_{n-1}}(k_\bot),
\ \ \ |k_\bot| \equiv \sqrt{k_\bot^2}.
\label{Conv_property}
\ee

By iterating
(\ref{Recurrence_rel})
one can work out the explicit expression for the operator
$X^{(n)}_{\mu_1 \ldots \mu_{n}}(k_\bot)$:
\be
&&
X_{\mu_1 \ldots \mu_n}^{(n)}(k_\bot)
\nonumber \\ &&
= \alpha(n) \left[ k^\bot_{\mu_1} \ldots k^\bot_{\mu_n}
- \frac{|k_\bot|^2}{2n-1} \left(
\sum_{i,j=1 \atop i<j}^n
g^\bot_{\mu_i \mu_j}
k^\bot_{\mu_1}
\ldots
{\underset{\land}{k^\bot_{\mu_i}}}
\ldots
{\underset{\land}{k^\bot_{\mu_j}}}
\ldots
k^\bot_{\mu_n} \right) \right.-
\frac{|k_\bot|^4}{(2n-1)(2n-3)}
\nonumber \\ &&
\times
\left. \left(
\sum_{i,j=1 \atop i<j}^n \sum_{k,l=1 \atop k<l \, k,l \ne i,j}^n
g^\bot_{\mu_i \mu_j} g^\bot_{\mu_k \mu_l}
k^\bot_{\mu_1}\ldots
{\underset{\land}{k^\bot_{\mu_i}}}
\ldots
{\underset{\land}{k^\bot_{\mu_j}}}
\ldots
{\underset{\land}{k^\bot_{\mu_k}}}
\ldots
{\underset{\land}{k^\bot_{\mu_l}}}
\ldots k^\bot_{\mu_n}
\right)+
\ldots
\right],
\label{IterSolution}
\ee
where  throughout this paper
${\underset{\land}{k^\bot_{\mu_i}}}$
denotes that the $i$-th entry
$k^\bot_{\mu_i}$
is omitted in the  $k^\bot_{\mu_1} \ldots k^\bot_{\mu_n}$ range:
$$k^\bot_{\mu_1}\ldots
{\underset{\land}{k^\bot_{\mu_i}}}
\ldots k^\bot_{\mu_n}
\equiv
k^\bot_{\mu_1}\ldots k^\bot_{\mu_{i-1}}k^\bot_{\mu_{i+1}}
\ldots k^\bot_{\mu_n};
$$
and
$\alpha(n)$
stands for the normalization factor.

Employing the convolution identity
(\ref{Conv_property})
together with the recurrence relation
(\ref{Recurrence_rel})
one can work out the normalization of the OAM-operators:
\be
X^{(n)}_{\mu_1 \ldots \mu_{n}}(k_\bot) X^{(n)}_{\mu_1 \ldots \mu_{n}}(k_\bot)=
\alpha(n) |k_\bot|^{2n}; \ \ \ \alpha(n)=\frac{(2n-1)!!}{n!}.
\label{Normalization_OAM_X}
\ee

One can check that the following generalization of the convolution relation (\ref{Conv_property})
is valid for $n \ge l$:
\be
X^{(n)}_{\mu_1 \ldots \mu_{n}}(k_\bot)
X^{(l)}_{\mu_1 \ldots \mu_{l}}(k_\bot)=\alpha(l) |k_\bot|^{2l} X^{(n-l)}_{\mu_1 \ldots \mu_{n-l}}(k_\bot).
\label{Conv_property_extra}
\ee
Note that for
$l=1$
one recovers the convolution relation
(\ref{Conv_property})
while
$l=n$
leads to the normalization condition
(\ref{Normalization_OAM_X}).

The convolution of
two spin-$n$ OAM-operators with the momenta
$k$
and
$q$
is given by
\be
X^{(n)}_{\mu_1 \ldots \mu_n}(k_\bot)
X^{(n)}_{\mu_1 \ldots \mu_n}(q_\bot)= \alpha(n) |q_\bot|^n |k_\bot|^n P_n(z);
\label{Conv_X_master}
\ee
where
$z= \frac{(q_\bot ,\, k_\bot)}{|q_\bot| |k_\bot|}$,
and
$P_n(z)$
are the Legendre polynomials.

\section{Differential technique for OAM-operators}
\label{Diff_op}
\mbox

In this section we adopt the covariant version of C.~Zemach's
O$(3)$ differential technique
\cite{Zemach:1968zz}
for the case of  OAM-operators
$X^{(n)}_{\mu_1 \ldots \mu_{n}}(k_\bot)$.
We consider the derivative operation
$D_\mu^{(n+1)}(k_\bot)$:
\be
D^{(n+1)}_\mu(k_\bot)
=
\frac{1}{n+1} \left( (2(n+1)+1) \frac{k^\bot_\mu}{|k_\bot|^2} -
\frac{k^\bot_\mu}{|k_\bot|} \frac{\partial}{\partial |k_\bot| } - \frac{\partial}{\partial k_\bot^\mu}\right).
\label{D_op_k}
\ee
We would like to show that
\be
D_\mu^{(n+1)}(k_\bot) \left\{ |k_\bot|^2 X^{(n)}_{\mu_1 \ldots \mu_n}(k_\bot) \right\}
=X^{(n+1)}_{\mu \mu_1 \ldots \mu_n}(k_\bot).
\label{Statement}
\ee
This derivative operation can be seen as
the ``inversion'' of the convolution formula
(\ref{Conv_property}).
It allows the iterative construction of the spin-$(n+1)$
OAM-operator from the spin-$n$ OAM-operator.

For our proof we employ the expression
(\ref{IterSolution})
for OAM-operators
obtained by iterating the recurrence expression 
(\ref{Recurrence_rel})
for
$X_{\mu_1 \ldots \mu_n}^{(n)}$.
This expansion is a polynomial in
$|k_\bot|^2$.
For given
$n$
the highest power is
$|k_\bot|^n$
for $n$-even and
$|k_\bot|^{n-1}$
for $n$-odd.
Let us consider the action of the derivative operator
(\ref{D_op_k})
on the OAM-operator
\be
D^{(n+1)}_\mu(k^\bot) |k_\bot|^2 X_{\mu_1 \ldots \mu_n}^{(n)}(k_\bot).
\ee

For example, consider $|k_\bot|^0$ and $|k_\bot|^2$ terms in (\ref{IterSolution}):
\be
&&
D^{(n+1)}_\mu(k_\bot) |k_\bot|^2 \alpha(n) k^\bot_{\mu_1} \ldots k^\bot_{\mu_n}
=
\frac{1}{n+1}(2n+1) \alpha(n) k^\bot_{\mu} k^\bot_{\mu_1} \ldots k^\bot_{\mu_n}
\nonumber \\ &&
-\frac{|k_\bot|^2}{n+1} \alpha(n) \sum_{i=1}^n g^{\bot}_{\mu \mu_i}
k^\bot_{\mu_1} \ldots k^\bot_{\mu_{i-1}} k^\bot_{\mu_{i+1}} \ldots k^\bot_{\mu_n}
\nonumber \\ &&
=\underbrace{\alpha(n+1) k^\bot_{\mu} k^\bot_{\mu_1} \ldots k^\bot_{\mu_n}}_{|k_\bot|^0 \ \ {\rm term \ \ for \ \ } X_{\mu \mu_1 \ldots \mu_n}^{(n+1)}(k_\bot)}
\underbrace{-\frac{|k_\bot|^2}{2n+1} \alpha(n+1) \sum_{i=1}^n g^{\bot}_{\mu \mu_i}
k^\bot_{\mu_1} \ldots
{\underset{\land}{k^\bot_{\mu_i}}}
\ldots k^\bot_{\mu_n}}_{{\rm part \ \ of \ \ }
|k_\bot|^2 \ \ {\rm term \ \ for \ \ } X_{\mu \mu_1 \ldots \mu_n}^{(n+1)}(k_\bot)};
\ee
\be
&&
-D^{(n+1)}_\mu(k_\bot) \frac{|k_\bot|^4}{2n-1} \alpha(n)
\left(
\sum_{i,j=1 \atop i<j}^n
g^\bot_{\mu_i \mu_j}
k^\bot_{\mu_1}
\ldots
{\underset{\land}{k^\bot_{\mu_i}}}
\ldots
{\underset{\land}{k^\bot_{\mu_j}}}
\ldots
k^\bot_{\mu_n} \right)
\nonumber \\ && =
-\alpha(n) \frac{1}{n+1} \frac{2(n+1)+1-4}{2n-1} |k_\bot|^2
\left(
\sum_{i,j=1 \atop i<j}^n
g^\bot_{\mu_i \mu_j}
k^\bot_{\mu}
k^\bot_{\mu_1} \ldots
{\underset{\land}{k^\bot_{\mu_i}}}
\ldots
{\underset{\land}{k^\bot_{\mu_j}}}
\ldots
k^\bot_{\mu_n} \right)
\nonumber \\ &&
+\alpha(n) \frac{1}{n+1} \frac{1}{2n-1} |k_\bot|^4 \left(
\sum_{i,j=1 \atop i<j}^n \sum_{k=1 \atop k \ne i,j}^n
g^\bot_{\mu \mu_k} g^\bot_{\mu_i \mu_j}
k^\bot_{\mu}
k^\bot_{\mu_1} \ldots
{\underset{\land}{k^\bot_{\mu_i}}}
\ldots
{\underset{\land}{k^\bot_{\mu_j}}}
\ldots
k^\bot_{\mu_n} \right)
\nonumber \\ &&
=-\alpha(n+1) \frac{|k_\bot|^2}{2n+1}
\left(
\sum_{i,j=1 \atop i<j}^n
g^\bot_{\mu_i \mu_j}
k^\bot_{\mu}
k^\bot_{\mu_1} \ldots
{\underset{\land}{k^\bot_{\mu_i}}}
\ldots
{\underset{\land}{k^\bot_{\mu_j}}}
\ldots
k^\bot_{\mu_n} \right)
\nonumber \\ &&
+\alpha(n+1)   \frac{|k_\bot|^4}{(2n+1)(2n-1)} \left(
\sum_{i,j=1 \atop i<j}^n \sum_{k=1 \atop k \ne i,j}^n
g^\bot_{\mu \mu_k} g^\bot_{\mu_i \mu_j}
k^\bot_{\mu}
k^\bot_{\mu_1}
\ldots
{\underset{\land}{k^\bot_{\mu_i}}}
\ldots
{\underset{\land}{k^\bot_{\mu_j}}}
\ldots
k^\bot_{\mu_n} \right);
\nonumber \\ &&
\ee
and analogously for all higher order terms.

Therefore, we conclude that the result of
action of the operator
(\ref{D_op_k})
on
$|k_\bot|^2 \times \big\{
O(|k_\bot|^m)-{\rm term \ \ in \ \ } X_{\mu_1 \ldots \mu_n}^{(n)}(k_\bot) \big\}$
($2 \le m < n$, $m$ - even)
contributes into the expansion
(\ref{IterSolution})
of
$X_{\mu \mu_1 \ldots \mu_n}^{(n+1)}(k_\bot)$
at orders
$O(|k_\bot|^m)$
and
$O(|k_\bot|^{m+1})$
with the 
proper coefficients%
\footnote{The highest order $O(|k_\bot|^n)$ term
for even $n$ is somewhat special. In this case there is a contribution only into
 $O(|k_\bot|^n)$ term coming from it. }. By combining the two contributions
at each order in
$|k_\bot|
$
we recover the complete result for
$X_{\mu \mu_1 \ldots \mu_n}^{(n+1)}(k_\bot)$
coinciding with that from the recurrence relation
(\ref{IterSolution}).
This finalizes the proof  of
eq.~(\ref{Statement}).

The technique based on the use of the derivative
operation (\ref{Statement}) is extremely convenient
for the calculation of convolutions of  OAM-operators
with several open indices.
This technique turns to be fully equivalent to the covariant
differential technique for the so-called contracted
projectors suggested by M.~Scadron in Ref.~\cite{Scadron:1968zz}
(see also Chapter I of Ref~\cite{Alfaro_red_book} and Appendix A of Ref.~\cite{SemenovTianShansky:2007hv}
for a short review of the contracted
projectors method).
In fact,
\be
X^{(n)}_{\alpha_1 \ldots \alpha_n}(q_\bot)
X^{(n)}_{\alpha_1 \ldots \alpha_n}(k_\bot)=
{\cal P}^n(q,p,P),
\ee
where
${\cal P}^n(q,p,P)$
stands for the contracted
projectors (the numerator of the bosonic spin sum contracted with
the initial and final relative momenta):
\be
{\cal P}^n(q,p,P) \equiv \alpha(n)^2 q_{\mu_1} \ldots q_{\mu_n}
O_{\mu_1 \ldots \mu_n}^{\nu_1 \ldots \nu_n} k^{\nu_1} \ldots k^{\nu_n },
\ee
where
$O_{\mu_1 \ldots \mu_n}^{\nu_1 \ldots \nu_n}$
stands for the bosonic projection operator%
\footnote{For the explicit form and properties of the
bosonic projection operator
$O_{\mu_1 \ldots \mu_n}^{\nu_1 \ldots \nu_n}$
see {\it e.g.} Sec.3.2 of Ref.~\cite{Anisovich:2006bc}.}, that
projects an arbitrary rank-$n$ Lorentz tensor into a
tensor, which satisfies the conditions
(\ref{Prop_sym})-(\ref{Prop_trace}).

For example, let us consider the convolution of OAM-operators
with one open index.  It is straightforward to check that
\be
&&
X^{(n+1)}_{\mu \alpha_1 \ldots \alpha_n}(q_\bot)
X^{(n)}_{\alpha_1 \ldots \alpha_n}(k_\bot)=
D^{(n+1)}_\mu(q_\bot) \Big[ \alpha(n) |q_\bot|^{n+2} |k_\bot|^n P_n(z) \Big]
\nonumber \\ &&
=\frac{\alpha(n)}{(n+1)}  |q_\bot|^{n+2} |k_\bot|^n
\sum_{N=0}^1 C_\mu^{(N; \, 1,0)}(n;\,q_\bot,k_\bot) P^{(N)}_n(z),
\ee
where by
$P^{(N)}_n(z)$
we denote the $N$-th derivative of the $n$-th Legendre
polynomial and
\be
C^{(0; \, 1,0)}_\mu(n;\,q_\bot,k_\bot)=(n+1)\frac{q^\bot_\mu}{|q_\bot|^2};
\ \ \
C^{(1; \, 1,0)}_\mu(n;\,q_\bot,k_\bot)=-\frac{dz}{d q_\bot^\mu}.
\label{TS10}
\ee
Note that the tensor structures
(\ref{TS10})
possess the simple convolution
properties
\be
q_\bot^\mu C^{(0; \, 1,0)}_\mu (n;\, q_\bot,k_\bot)=(n+1);
\ \ \
q_\bot^\mu C^{(1; \, 1,0)}_\mu(n;\, q_\bot,k_\bot)=0.
\ee
This ensures the validity of the identity
\be
q_\bot^\mu
X^{(n+1)}_{\mu \alpha_1 \ldots \alpha_n}(q_\bot)
X^{(n)}_{\alpha_1 \ldots \alpha_n}(k_\bot)=
|q_\bot|^2
X^{(n)}_{\alpha_1 \ldots \alpha_n}(q_\bot)
X^{(n)}_{\alpha_1 \ldots \alpha_n}(k_\bot),
\ee
that is the consequence of
(\ref{Conv_property}).

Employing the explicit expression
(\ref{dzdqmu})
for
$\frac{dz}{d q_\bot^\mu}$
together with the well known recurrence relation for the derivative of the
Legendre polynomial
$$
P'_{n+1}(z)=z P'_n(z)+(n+1) P_n(z)
$$
we check that we indeed recover the eq.~(B.1) of Appendix B of
Ref.~\cite{Anisovich:2004zz}:
\be
&&
D^{(n+1)}_\mu(q_\bot) \Big[ \alpha(n) |q_\bot|^{n+2} |k_\bot|^n P_n(z) \Big]
\nonumber \\  &&
= \frac{\alpha(n)}{(n+1)}  |q_\bot|^{n+1} |k_\bot|^n
\Big\{
-\frac{k^\bot_\mu}{|k_\bot|} P'_n(z)+\frac{q^\bot_\mu}{|q_\bot|} P'_{n+1}(z)
\Big\}.
\ee

The application of this technique for the more involved cases
is presented in the Appendix~\ref{App_A}.

\pagebreak

\section{Some generalization of OAM-operators}
\mbox

Formally the OAM-operators
$X_{\mu_1 \ldots \mu_n}^{(n)}(k_\bot)$
are the most general
spin-$n$ operators satisfying the list of requirements
(\ref{Prop_sym})-- (\ref{Prop_trace})
constructed from a sole vector $k$.
This can be most clearly seen from the formula employing the
bosonic projection operator:
\be
 k_{\mu_1} \ldots k_{\mu_n}
O^{\mu_1 \ldots \mu_n}_{\nu_1 \ldots \nu_n}
=\frac{1}{\alpha(n)} X_{\nu_1 \ldots \nu_n}^{(n)}(k_\bot).
\ee

A natural generalization, which was already considered within the
non-covariant formalism of C.~Zemach
\cite{Zemach:1968zz},
are the operators constructed out of two, three and more independent
vectors. For example, let us consider the spin-$n$ operators
$X_{\nu_1 \ldots \nu_n}^{(n,l)}(k_\bot, q_\bot)$
constructed out
$n-l$ entries of the four-vector $k$ and  $l$  entries of the four-vector $q$ with 
$0 \le l \le n$:
\be
\underbrace{k_{\mu_1} \ldots k_{\mu_{n-l}}}_{n-l \ \ {\rm entries}}
 \underbrace{q_{\mu_{n-l+1}} \ldots  q_{\mu_n}}_{l \ \ {\rm entries}}
O^{\mu_1 \ldots \mu_n}_{\nu_1 \ldots \nu_n}
\equiv \frac{1}{\alpha(n)} X_{\nu_1 \ldots \nu_n}^{(n,l)}(k_\bot, q_\bot).
\ee

One can check that
\be
X_{\mu_1 \ldots \mu_n}^{(n,l)}(k_\bot, q_\bot)=
\left[ \prod_{i=0}^{l-1} \frac{1}{n-i} \left( q^\mu_\bot \frac{d}{dk_\bot^\mu} \right)
\right]
X_{\mu_1 \ldots \mu_n}^{(n)}(k_\bot).
\ee
Also it is straightforward to verify that for 
$l \le n$
\be
X^{(n)}_{\mu_1 \ldots \mu_{n}}(k_\bot)
X_{\mu_1 \ldots \mu_n}^{(n,l)}(k_\bot, q_\bot)
= \alpha(n) |k_\bot|^{2n-l} q_\bot^l P_l(z).
\label{Conv_new}
\ee

Obviously for $l=n$ we get
\be
X_{\mu_1 \ldots \mu_n}^{(n,l=n)}(k_\bot, q_\bot)=X_{\mu_1 \ldots \mu_n}^{(n)}(q_\bot)
\ee
and we recover
(\ref{Conv_X_master})
from
(\ref{Conv_new}).

One also can work out the following equation
\be
(2n+1) z X^{(n)}_{\mu_1 \ldots \mu_{n}}(k_\bot)=
n
\frac{|k_\bot|}{|q_\bot|}
X^{(n,1)}_{\mu_1 \ldots \mu_{n}}(k_\bot,q_\bot)+(n+1)
\frac{q_\bot^\mu}{|k_\bot||q_\bot|}
X^{(n+1)}_{\mu \mu_1 \ldots \mu_{n}}(k_\bot).
\label{New_rel_X}
\ee
Let us sketch the proof of
eq.~(\ref{New_rel_X}).
To get the first relation for the
coefficients of the two tensors in the r.h.s. of eq.~(\ref{New_rel_X})
one
has to set $q_\bot=p_\bot$.
To get  the second relation for the coefficients
one has to contract (\ref{New_rel_X})
with
$X^{(n)}_{\mu_1 \ldots \mu_{n}}(k_\bot)$
and employ
(\ref{Normalization_OAM_X})
together with
(\ref{Conv_new})
for $l=1$
and
(\ref{Conv_property_extra}).

Now, by contracting
(\ref{New_rel_X})
with
$X^{(n)}_{\mu_1 \ldots \mu_{n}}(k_\bot)$,
one can see explicitly the property of the OAM-operators
related to the well-known recurrence relation for the
Legendre polynomials:
\be
&&
(2n+1) z \underbrace{X^{(n)}_{\mu_1 \ldots \mu_{n}}(k_\bot)
X^{(n)}_{\mu_1 \ldots \mu_{n}}(q_\bot)}_{\alpha(n) |k_\bot|^n |q_\bot|^n P_n(z)}=
n
\frac{|k_\bot|}{|q_\bot|}
\underbrace{X^{(n,1)}_{\mu_1 \ldots \mu_{n}}(k_\bot,q_\bot) X^{(n)}_{\mu_1 \ldots \mu_{n}}(q_\bot)}_{\alpha(n) |k_\bot|^{n-1} |q_\bot|  P_{n-1}(z)}
\nonumber \\ &&
+
(n+1)
\frac{1}{|k_\bot||q_\bot|}
\underbrace{X^{(n+1)}_{\mu \mu_1 \ldots \mu_{n}}(k_\bot) q_\bot^\mu
X^{(n)}_{\mu_1 \ldots \mu_{n}}(q_\bot)}_{\alpha(n+1) \frac{n+1}{2n+1}
|k_\bot|^{n+1} |k_\bot|^{n+1} P_{n+1}(z) }.
\label{Prove_recrel}
\ee
Obviously this is nothing but the familiar  relation
for the Legendre polynomials
\be
(2n+1) z P_n(z)= n P_n(z)+(n+1) P_{n+1}(z)
\ee
known as  Bonnet's recursion formula.
Note that in the last term in
(\ref{Prove_recrel})
we used the identity
\be
O^{\mu \mu_1 \ldots \mu_n}_{\nu \nu_1 \ldots \nu_n}
q_\mu  X^{(n)}_{\mu_1 \ldots \mu_{n}}(q_\bot)= \frac{n+1}{2n+1}
X^{(n+1)}_{\nu \nu_1 \ldots \nu_{n}}(q_\bot),
\ee
that can be easily established from
eq.~(\ref{IterSolution}).

The generalized OAM-operators introduced in this Section
can be useful for the development of covariant and
fully
electromagnetically gauge invariant description
of hadron photo- and electroproduction reactions

\section{Conclusions}
\label{Sec_Concl}
\mbox

Convolutions of OAM-operators with several open Lorentz indices
occur in the description of photo- and electroproduction
of mesons off nucleons within the covariant 
approach for construction of
partial wave amplitudes.
The differential technique developed in the present paper considerably
simplifies the calculation of such OAM-operator convolutions.
These findings will greatly help the development of the fully
electromagnetically gauge invariant description
of hadron photo- and electroproduction reactions within the
covariant OAM-operator expansion approach.

\section*{Acknowledgements}
\mbox

K.S. is grateful to V. Vereshagin for very instructive discussions
on the method of contracted projectors. The paper was
supported by grant RSF 16-12-10267.

\setcounter{section}{0}
\setcounter{equation}{0}
\renewcommand{\thesection}{\Alph{section}}
\renewcommand{\theequation}{\thesection\arabic{equation}}

\section{Useful convolutions of orbital-angular-momentum-operators}
\label{App_A}
\mbox

In this Appendix we present some useful convolutions of OAM-operators with several open indices occurring in the
calculation of baryon electroproduction amplitudes with the use
of the set of effective vertices worked out in \cite{Anisovich:2016hmt}.
This can be easily done with the help of the derivative operation (\ref{Statement}).

\subsection{Case of $3$ open indices}
\mbox

To derive the explicit expression for the convolution of OAM-operators
with $3$ open indices we apply the
$D_{\nu}^{(n+1)}(k_\bot)$
operator to the expression for the convolution of OAM-operators
with $2$ open indices given by
eq.~(B.3) of Ref.~\cite{Anisovich:2004zz}:
$|k_\bot|^2 X^{(n+2)}_{\mu_1 \mu_2 \alpha_1 \ldots \alpha_{n}}(q_\bot) X^{(n)}_{ \alpha_1 \ldots \alpha_{n}}(k_\bot)$.

We get
\be
&&
D_{\nu}^{(n+1)}(k_\bot) \Big[ |k_\bot|^2 \underbrace{X^{(n+2)}_{\mu_1 \mu_2 \alpha_1 \ldots \alpha_{n}}(q_\bot) X^{(n)}_{ \alpha_1 \ldots \alpha_{n}}(k_\bot)}_{\rm Expressed \ \ by \ \ eq.~(B.3) \ \ of \ \ Ref.~\cite{Anisovich:2004zz}} \Big] \nonumber \\ &&
=\frac{\alpha(n)}{(n+2)(n+1)^2} |q_\bot|^{n+4} |k_\bot|^{n+2}
\sum_{N=0}^3 C^{(N; \,2,1)  }_{\mu_1 \mu_2 \nu}(n; \, q_\bot,k_\bot) P_n^{(N)}(z),
\label{3IndexX}
\ee
where
\be
&&
C^{(0; \, 2,1)  }_{\mu_1 \mu_2 \nu}(n; \, q_\bot,k_\bot)=(n+1)^2 \left((n+3)\frac{q^\bot_{\mu_1}q^\bot_{\mu_2}}{|q_\bot|^4}- \frac{g^\bot_{\mu_1 \mu_2}}{|q_\bot|^2} \right) \frac{k^\bot_\nu}{|k_\bot|^2};
\nonumber \\ &&
C^{(1; \, 2,1)  }_{\mu_1 \mu_2 \nu}(n; \, q_\bot,k_\bot)=
-\frac{d^3 z}{d q^{\mu_1}_\bot d q^{\mu_2}_\bot d k^{\nu}_\bot }
\nonumber \\ &&
+
(n+1) \frac{d^2 z}{d q^{\mu_1}_\bot d q^{\mu_2}_\bot}  \frac{k^\bot_\nu}{|k_\bot|^2}
+
(n+1) \frac{d^2 z}{d q^{\mu_1}_\bot d k^{\nu}_\bot}  \frac{q^\bot_{\mu_2}}{|q_\bot|^2}
+
(n+1) \frac{d^2 z}{d q^{\mu_2}_\bot d k^{\nu}_\bot}  \frac{q^\bot_{\mu_1}}{|q_\bot|^2}
\nonumber \\ &&
-(n+1)^2 \frac{d z}{d q^{\mu_1}_\bot}  \frac{q^\bot_{\mu_2} k^\bot_\nu}{|q_\bot|^2|k_\bot|^2}
-(n+1)^2 \frac{d z}{d q^{\mu_2}_\bot}  \frac{q^\bot_{\mu_1} k^\bot_\nu}{|q_\bot|^2|k_\bot|^2}
\nonumber \\ &&
-
(n+1) \left((n+3)\frac{q^\bot_{\mu_1}q^\bot_{\mu_2}}{|q_\bot|^4}- \frac{g^\bot_{\mu_1 \mu_2}}{|q_\bot|^2} \right) \frac{d z}{d k_\bot^\nu};
\nonumber \\ &&
C^{(2; \, 2,1)  }_{\mu_1 \mu_2 \nu}(n; \, q_\bot,k_\bot)=
-\frac{d^2 z}{d q^{\mu_1}d q^{\mu_2}}\frac{d z}{d k^{\nu}}
-\frac{d^2 z}{d q^{\mu_1}d k^{\nu}}\frac{d z}{d q^{\mu_2}}
-\frac{d^2 z}{d q^{\mu_2}d k^{\nu}}\frac{d z}{d q^{\mu_1}}
\nonumber \\ &&
+(n+1) \frac{d z}{d q^{\mu_1}_\bot} \frac{d z}{d q^{\mu_2}_\bot} \frac{k^\bot_\nu}{|k_\bot|^2}
+(n+1) \frac{d z}{d q^{\mu_1}_\bot} \frac{d z}{d k^{\nu}_\bot} \frac{q^\bot_{\mu_{2}}}{|q_\bot|^2}
+(n+1) \frac{d z}{d q^{\mu_2}_\bot} \frac{d z}{d k^{\nu}_\bot} \frac{q^\bot_{\mu_{1}}}{|q_\bot|^2};
\nonumber \\ &&
C^{(3; \, 2,1)  }_{\mu_1 \mu_2 \nu}(n; \, q_\bot,k_\bot)= -\frac{d z}{d q^{\mu_1}_\bot} \frac{d z}{d q^{\mu_2}_\bot}
\frac{d z}{d k^{\nu}_\bot}.
\ee
Note that all the tensor structures
$C^{(N; \, 2,1)  }_{\mu_1 \mu_2 \nu}(n; \, q_\bot,k_\bot)$
are symmetric under the permutation
$\mu_1 \leftrightarrow \mu_2$,
and possess the simple convolution properties:
\be
&&
k_\bot^\nu C^{(N; \, 2,1)  }_{\mu_1 \mu_2 \nu}(n; \, q_\bot,k_\bot)
=(n+1) C^{(N; \, 2,0)  }_{\mu_1 \mu_2 }(n; \, q_\bot,k_\bot) \ \ \ {\rm for} \ \ N=0,1,2;
\nonumber \\ &&
k_\bot^\nu C^{(3; \, 2,1)  }_{\mu_1 \mu_2 \nu}(n; \, q_\bot,k_\bot)=0;
\nonumber \\ &&
q_\bot^{\mu_2} C^{(N; \, 2,1)  }_{\mu_1 \mu_2 \nu}(n; \, q_\bot,k_\bot)
=(n+2) C^{(N; \, 1,1)  }_{\mu_1  \nu}(n; \, q_\bot,k_\bot) \ \ \ {\rm for} \ \ N=0,1,2;
\nonumber \\ &&
q_\bot^{\mu_2} C^{(3; \, 2,1)  }_{\mu_1 \mu_2 \nu}(n; \, q_\bot,k_\bot)=0\,.
\label{ConVrelC21}
\ee

\pagebreak

\subsection{Case of $4$ open indices}
\mbox

Finally, for the previously poorly known convolution with $4$ open indices
\be
X^{(n+2)}_{\mu_1 \mu_2 \alpha_1 \ldots \alpha_{n}}(q_\bot)
X^{(n+2)}_{\nu_1 \nu_2 \alpha_1 \ldots \alpha_{n}}(k_\bot)
\ee
we can write the following formula
\be
&&
X^{(n+2)}_{\mu_1 \mu_2 \alpha_1 \ldots \alpha_{n}}(q_\bot)
X^{(n+2)}_{\nu_1 \nu_2 \alpha_1 \ldots \alpha_{n}}(k_\bot) \nonumber \\ &&
=
D_{{\nu_2}}^{(n+2)}(k_\bot) \Big[ |k_\bot|^2 \underbrace{X^{(n+2)}_{\mu_1 \mu_2 \alpha_1 \ldots \alpha_{n}}(q_\bot) X^{(n+1)}_{\nu_1 \alpha_1 \ldots \alpha_{n}}(k_\bot)}_{{\rm Expressed \ \ by \ \ eq.~(\ref{3IndexX})}} \Big]
\label{4IndexX}
\ee
Performing explicitly the derivative operation we get
\be
&&
X^{(n+2)}_{\mu_1 \mu_2 \alpha_1 \ldots \alpha_{n}}(q_\bot)
X^{(n+2)}_{\nu_1 \nu_2 \alpha_1 \ldots \alpha_{n}}(k_\bot)
\nonumber \\ &&=
\frac{\alpha(n)}{(n+2)^2(n+1)^2} |q_\bot|^{n+4} |k_\bot|^{n+4}
\sum_{N=0}^4 C^{(N; \, 2,2)}_{\mu_1 \mu_2 \nu_1 \nu_2}(n;\, q_\bot,k_\bot) P_n^{(N)}(z),
\ee
where
\be
&&
C^{(0; \, 2,2)}_{\mu_1 \mu_2 \nu_1 \nu_2}(n;\, q_\bot,k_\bot)=
(n+1)^2 \left((n+3)\frac{q^\bot_{\mu_1}q^\bot_{\mu_2}}{|q_\bot|^4}- \frac{g^\bot_{\mu_1 \mu_2}}{|q_\bot|^2} \right)
\left((n+3)\frac{k^\bot_{\nu_1}k^\bot_{\nu_2}}{|k_\bot|^4}- \frac{g^\bot_{\nu_1 \nu_2}}{|k_\bot|^2} \right);
\nonumber 
\ee
\be
&&
C^{(1; \, 2,2)}_{\mu_1 \mu_2 \nu_1 \nu_2}(n;\, q_\bot,k_\bot)=
\frac{d^4z}{d q_\bot^{\mu_1} d q_\bot^{\mu_2} d k_\bot^{\nu_1} d k_\bot^{\nu_2}}
\nonumber \\ &&
-(n+1) \left( \frac{k^\bot_{\nu_1}}{|k_\bot|^2}
\frac{d^3 z}{d q_\bot^{\mu_1} d q_\bot^{\mu_2} d k_\bot^{\nu_2}}+ \{\nu_1 \leftrightarrow \nu_2 \} \right)
-
(n+1) \left( \frac{q^\bot_{\mu_1}}{|q_\bot|^2}
\frac{d^3 z}{d k_\bot^{\nu_1} d k_\bot^{\nu_2} d q_\bot^{\mu_2}}+ \{\mu_1 \leftrightarrow \mu_2 \} \right)
\nonumber \\ &&
+(n+1)\left( (n+3) \frac{k^\bot_{\nu_1} k^\bot_{\nu_2}}{|k_\bot|^4} -
\frac{g^\bot_{\nu_1 \nu_2}}{|k_\bot|^2}
\right) \frac{d^2 z}{d q_\bot^{\mu_1} d q_\bot^{\mu_2}}
+(n+1)\left( (n+3) \frac{q^\bot_{\mu_1} q^\bot_{\mu_2}}{|q_\bot|^4} -
\frac{g^\bot_{\mu_1 \mu_2}}{|q_\bot|^2}
\right) \frac{d^2 z}{d k_\bot^{\nu_1} d k_\bot^{\nu_2}}
\nonumber \\ &&
+(n+1)^2
\left( \frac{q^\bot_{\mu_1} k^\bot_{\nu_1}}{|q_\bot|^2 |k_\bot|^2}
\frac{d^2z}{d q_\bot^{\mu_2} d k_\bot^{\nu_2}}+ \{ \nu_1 \leftrightarrow \nu_2 \}
\right)
+(n+1)^2
\left( \frac{q^\bot_{\mu_2} k^\bot_{\nu_1}}{|q_\bot|^2 |k_\bot|^2}
\frac{d^2z}{d q_\bot^{\mu_1} d k_\bot^{\nu_2}}+ \{ \nu_1 \leftrightarrow \nu_2 \}
\right)
\nonumber \\ &&
-(n+1)^2 \left( \frac{dz}{d q_\bot^{\mu_1}}
\left(
(n+3)\frac{q^\bot_{\mu_2} k^\bot_{\nu_1}k^\bot_{\nu_2}}{|q_\bot|^2 |k_\bot|^4}-
\frac{q^\bot_{\mu_2} g^\bot_{\nu_1 \nu_2}}{|q_\bot|^2 |k_\bot|^2}
\right)
+ \{ \mu_1 \leftrightarrow \mu_2 \}
\right)
\nonumber \\ &&
-(n+1)^2 \left( \frac{dz}{d k_\bot^{\nu_1}}
\left(
(n+3)\frac{q^\bot_{\mu_1} q^\bot_{\mu_2} k^\bot_{\nu_2}}{|q_\bot|^4 |k_\bot|^2}-
\frac{ k^\bot_{\nu_2} g^\bot_{\mu_1 \mu_2 }}{|q_\bot|^2 |k_\bot|^2}
\right)
+ \{ \nu_1 \leftrightarrow \nu_2 \}
\right); \nonumber
\ee
\be
 &&
C^{(2; \, 2,2)}_{\mu_1 \mu_2 \nu_1 \nu_2}(n;\, q_\bot,k_\bot)
\nonumber \\ &&
=
\left(\frac{d^3 z}{d q_\bot^{\mu_1} d q_\bot^{\mu_2} d k_\bot^{\nu_1}} \frac{dz}{d k_\bot^{\nu_2}}+ \{\nu_1 \leftrightarrow \nu_2 \} \right)+
\left(\frac{d^3 z}{d q_\bot^{\mu_1} d k_\bot^{\nu_1} d k_\bot^{\nu_2}} \frac{dz}{d q_\bot^{\mu_2}}+ \{\mu_1 \leftrightarrow \mu_2 \} \right)
\nonumber \\ &&
+ \frac{d^2 z}{d q_\bot^{\mu_1} d q_\bot^{\mu_2}} \frac{d^2 z}{d k_\bot^{\nu_1} d k_\bot^{\nu_2}}
+ \frac{d^2 z}{d q_\bot^{\mu_1} d k_\bot^{\nu_1}} \frac{d^2 z}{d q_\bot^{\mu_2} d k_\bot^{\nu_2}}
+ \frac{d^2 z}{d q_\bot^{\mu_1} d k_\bot^{\nu_2}} \frac{d^2 z}{d q_\bot^{\mu_2} d k_\bot^{\nu_1}}
\nonumber \\ &&
-(n+1) \left(
\frac{d^2 z}{d q_\bot^{\mu_1} d q_\bot^{\mu_2}}
\left( \frac{k^\bot_{\nu_1}}{|k_\bot|^2} \frac{d z}{d k_\bot^{\nu_2}}
+ \frac{k^\bot_{\nu_2}}{|k_\bot|^2} \frac{d z}{d k_\bot^{\nu_1}}
\right)
+\{\mu \leftrightarrow \nu; \; q_\bot \leftrightarrow k_\bot \}\right)
\nonumber \\ &&
-(n+1) \left(
\frac{d^2 z}{d q_\bot^{\mu_1} d k_\bot^{\nu_2}}
\left( \frac{k^\bot_{\nu_1}}{|k_\bot|^2} \frac{d z}{d q_\bot^{\mu_2}}
+ \frac{q^\bot_{\mu_2}}{|k_\bot|^2} \frac{d z}{d k_\bot^{\nu_1}}
\right)
+\{\mu_1 \leftrightarrow \mu_2 \}\right)
\nonumber \\ &&
-(n+1) \left(
\frac{d^2 z}{d q_\bot^{\mu_2} d k_\bot^{\nu_1}}
\left( \frac{k^\bot_{\nu_2}}{|k_\bot|^2} \frac{d z}{d q_\bot^{\mu_1}}
+ \frac{q^\bot_{\mu_1}}{|k_\bot|^2} \frac{d z}{d k_\bot^{\nu_2}}
\right)
+\{\nu_1 \leftrightarrow \nu_2 \}\right)
\nonumber \\ &&
+(n+1)^2
\left(
\frac{dz}{dq_\bot^{\mu_1}} \frac{dz}{dk_\bot^{\nu_1}}
\frac{q^\bot_{\mu_2} k^\bot_{\nu_2}}{|q_\bot|^2|k_\bot|^2}+
\frac{dz}{dq_\bot^{\mu_1}} \frac{dz}{dk_\bot^{\nu_2}}
\frac{q^\bot_{\mu_2} k^\bot_{\nu_1}}{|q_\bot|^2|k_\bot|^2}
\right.
\nonumber \\ &&
\left.
+
\frac{dz}{dq_\bot^{\mu_2}} \frac{dz}{dk_\bot^{\nu_1}}
\frac{q^\bot_{\mu_1} k^\bot_{\nu_2}}{|q_\bot|^2|k_\bot|^2}+
\frac{dz}{dq_\bot^{\mu_2}} \frac{dz}{dk_\bot^{\nu_2}}
\frac{q^\bot_{\mu_1} k^\bot_{\nu_1}}{|q_\bot|^2|k_\bot|^2}
\right)
\nonumber \\ &&
+\frac{dz}{d q_\bot^{\mu_1}}
\frac{dz}{d q_\bot^{\mu_2}}
(n+1)\left((n+3) \frac{k^\bot_{\nu_1}k^\bot_{\nu_2}}{|k_\bot|^4} -\frac{g^\bot_{\nu_1 \nu_2}}{|k_\bot|^2} \right)
+\frac{dz}{d k_\bot^{\nu_1}}
\frac{dz}{d k_\bot^{\nu_2}}
(n+1)\left((n+3) \frac{q^\bot_{\mu_1}q^\bot_{\mu_2}}{|q_\bot|^4} -\frac{g^\bot_{\nu_1 \nu_2}}{|q_\bot|^2} \right);
\nonumber
\ee

\be
 &&
C^{(3; \, 2,2)}_{\mu_1 \mu_2 \nu_1 \nu_2}(n;\, q_\bot,k_\bot)
=
\frac{d^2 z}{d q_\bot^{\mu_1} d q_\bot^{\mu_2}} \frac{dz}{d k_\bot^{\nu_1}}
\frac{dz}{d k_\bot^{\nu_2}}
+
\frac{d^2 z}{d k_\bot^{\nu_1} d k_\bot^{\nu_2}} \frac{dz}{d q_\bot^{\mu_1}}
\frac{dz}{d q_\bot^{\mu_2}}
\nonumber \\ &&
+\left( \frac{d^2 z}{d q_\bot^{\mu_1} d k_\bot^{\nu_1}}
\frac{dz}{d q_\bot^{\mu_2}} \frac{dz}{d k_\bot^{\nu_2}}+
3 \; {\rm permutations}
\right)
\nonumber \\ &&
-(n+1) \left(
\frac{dz}{d q_\bot^{\mu_1}} \frac{dz}{d q_\bot^{\mu_2}}
\frac{dz}{d k_\bot^{\nu_1}} \frac{k^\bot_{\nu_2}}{|k_\bot|^2}
+
\frac{dz}{d q_\bot^{\mu_1}} \frac{dz}{d q_\bot^{\mu_2}}
\frac{dz}{d k_\bot^{\nu_2}} \frac{k^\bot_{\nu_1}}{|k_\bot|^2}
\right.
\nonumber \\ &&
\left.
+
\frac{dz}{d q_\bot^{\mu_1}} \frac{dz}{d k_\bot^{\nu_1}}
\frac{dz}{d k_\bot^{\nu_2}} \frac{q^\bot_{\mu_2}}{|q_\bot|^2}
+
\frac{dz}{d q_\bot^{\mu_2}} \frac{dz}{d k_\bot^{\nu_1}}
\frac{dz}{d k_\bot^{\nu_2}} \frac{q^\bot_{\mu_1}}{|q_\bot|^2}
\right);
\nonumber \\ &&
C^{(4; \, 2,2)}_{\mu_1 \mu_2 \nu_1 \nu_2}(n;\, q_\bot,k_\bot)=
\frac{d z}{d q^{\mu_1}_\bot} \frac{d z}{d q^{\mu_2}_\bot}
\frac{d z}{d k^{\nu_1}_\bot} \frac{d z}{d k^{\nu_2}_\bot}\,. \nonumber
\ee
Note that the coefficients
$C^{(N;\,2,2)}$
are symmetric under
$\mu_1 \leftrightarrow \mu_2$,
$\nu_1 \leftrightarrow \nu_2$
and
$q_\bot \leftrightarrow  k_\bot$
permutations.

These coefficient also possess the nice convolution properties. Namely,
\be
&&
k_\bot^{\nu_1} C^{(N;\,2,2)}_{\mu_1 \mu_2 \nu_1 \nu_2}(n;\, q_\bot,k_\bot)=(n+2)
C^{(N;\,2,1)}_{\mu_1 \mu_2 \nu_2}(n;\, q_\bot,k_\bot); \ \ \ N=0,\,1,\,2,\,3;
\nonumber \\ &&
k_\bot^{\nu_1} C^{(4;\,2,2)}_{\mu_1 \mu_2 \nu_1 \nu_2}(n;\, q_\bot,k_\bot)=0;
\nonumber \\ &&
k_\bot^{\nu_1} k_\bot^{\nu_2} C^{(N;\,2,2)}_{\mu_1 \mu_2 \nu_1 \nu_2}(n;\, q_\bot,k_\bot)=(n+2)(n+1)
C^{(N;\,2,0)}_{\mu_1 \mu_2}(n;\, q_\bot,k_\bot); \ \ \ N=0,\,1,\,2;
\nonumber \\ &&
k_\bot^{\nu_1} k_\bot^{\nu_2} C^{(3;\,2,2)}_{\mu_1 \mu_2 \nu_1 \nu_2}(n;\, q_\bot,k_\bot)=0;
\nonumber \\ &&
k_\bot^{\nu_1} k_\bot^{\nu_2} q_\bot^{\mu_2} C^{(N;\,2,2)}_{\mu_1 \mu_2 \nu_1 \nu_2}(n;\, q_\bot,k_\bot)=(n+2)^2(n+1)
C^{(N;\,1,0)}_{\mu_1}(n;\, q_\bot,k_\bot); \ \ \ N=0,\,1;
\nonumber \\ &&
k_\bot^{\nu_1} k_\bot^{\nu_2} q_\bot^{\mu_1} q_\bot^{\mu_2} C^{(0;\,2,2)}_{\mu_1 \mu_2 \nu_1 \nu_2}(n;\, q_\bot,k_\bot)=(n+2)^2(n+1)^2.
\label{ConVrelC22}
\ee

\section{Miscellaneous}
\label{App_B}
\mbox

In this Appendix we summarize the explicit expressions for the
set of tensor structures occurring in the 
tensor coefficients
$C^{(N;\,i,j)}_{\mu_1 \ldots \mu_i \nu_1 \ldots \nu_j}$
within the convolutions of angular
momentum operators summarized in Appendix~\ref{App_A}.

\be
\frac{dz}{dq_{\bot}^{\mu }}= \frac{k^\bot_\mu}{|q_\bot||k_\bot|}- z \frac{q^\bot_\mu}{|q_\bot|^2}.
\label{dzdqmu}
\ee
Note that
$$
\frac{dz}{dq_{\bot}^{\mu }} q^{\bot}_{ \mu }=0.
$$

\be
\frac{d^2z}{dq_{\bot}^{\mu_1 } dq_{\bot}^{\mu_2 }}=
3 z \frac{q^\bot_{\mu_1}q^\bot_{\mu_2}}{|q_\bot|^4}-
\frac{q^\bot_{\mu_1} k^\bot_{\mu_2}+q^\bot_{\mu_2} k^\bot_{\mu_1}}{|q_\bot|^3|k_\bot|}-
z \frac{g^\bot_{\mu_1 \mu_2}}{|q_\bot|^2}.
\ee
The tensor is symmetric under $\mu_1 \leftrightarrow \mu_2$
permutation and satisfies
$$
\frac{d^2z}{dq_{\bot}^{\mu_1 } dq_{\bot}^{\mu_2 }} q_{\bot}^{\mu_2 }=
- \frac{dz}{dq_{\bot}^{\mu_1 }}.
$$

\be
\frac{d^2z}{dq_{\bot}^{\mu } dk_{\bot}^{\nu }}=
 z \frac{q^\bot_{\mu}k^\bot_{\nu}}{|q_\bot|^2|k_\bot|^2}-
\frac{q^\bot_{\mu} q^\bot_{\nu}}{|q_\bot|^3|k_\bot|}-
\frac{k^\bot_{\mu} k^\bot_{\nu}}{|q_\bot||k_\bot|^3}
+
 \frac{g^\bot_{\mu \nu}}{|q_\bot||k_\bot|}.
\ee
This tensor is obviously symmetric under
$q_{\bot } \leftrightarrow  k_{\bot }$ interchange and
satisfies
$$
\frac{d^2z}{dq_{\bot}^{\mu } dk_{\bot}^{\nu }} k_{\bot}^{\nu }=0; \ \ \
\frac{d^2z}{dq_{\bot}^{\mu } dk_{\bot}^{\nu }} q_{\bot}^{\mu }=0;
$$

\be
&&
\frac{d^3z}{dq_{\bot}^{\mu_1 } dq_{\bot}^{\mu_2 } dk_{\bot}^{\nu }}=
3  \frac{q^\bot_{\mu_1}q^\bot_{\mu_2} q^\bot_\nu}{|q_\bot|^5 |k_\bot|}
-3z \frac{q^\bot_{\mu_1}q^\bot_{\mu_2} k^\bot_\nu}{|q_\bot|^4 |k_\bot|^2}
+
\frac{q^\bot_{\mu_2} k^\bot_{\mu_1} k^\bot_\nu+ q^\bot_{\mu_1} k^\bot_{\mu_2} k^\bot_\nu}{|q_\bot|^3 |k_\bot|^3} \\ &&
- \frac{g^\bot_{\mu_1 \nu} q^\bot_{\mu_2} + g^\bot_{\mu_2 \nu} q^\bot_{\mu_1}}{|q_\bot|^3 |k_\bot|}-
\frac{g^\bot_{\mu_1 \mu_2} q^\bot_\nu}{|q_\bot|^3 |k_\bot|}+
z \frac{g^\bot_{\mu_1 \mu_2} k^\bot_\nu}{|q_\bot|^2 |k_\bot|^2}.
\ee

The tensor is symmetric under $\mu_1 \leftrightarrow \mu_2$
permutation and satisfies
\be
\frac{d^3z}{dq_{\bot}^{\mu_1 } dq_{\bot}^{\mu_2 } dk_{\bot}^{\nu }}
k_{\bot  }^\nu=0;
\ee

\be
\frac{d^3z}{dq_{\bot}^{\mu_1 } dq_{\bot}^{\mu_2 } dk_{\bot}^{\nu }}
q_{\bot }^{\mu_2}=-
\frac{d^2z}{dq_{\bot}^{\mu_1 } dk_{\bot}^{\nu }}.
\ee

Finally,
\be
&&
\frac{d^4z}{dq_{\bot}^{\mu_1 } dq_{\bot}^{\mu_2 } dk_{\bot}^{\nu_1 } dk_{\bot}^{\nu_2 }}=
9z \frac{q^\bot_{\mu_1} q^\bot_{\mu_2} k^\bot_{\nu_2} k^\bot_{\nu_2}}{|q_\bot|^4 |k_\bot|^4}
-3\frac{
k^\bot_{\mu_1} q^\bot_{\mu_2} k^\bot_{\nu_1} k^\bot_{\nu_2}+
q^\bot_{\mu_1} k^\bot_{\mu_2} k^\bot_{\nu_1} k^\bot_{\nu_2}
}{|q_\bot|^3 |k_\bot|^5} \nonumber
\\ &&
-3\frac{
q^\bot_{\mu_1} q^\bot_{\mu_2} q^\bot_{\nu_1} k^\bot_{\nu_2}+
q^\bot_{\mu_1} q^\bot_{\mu_2} k^\bot_{\nu_1} q^\bot_{\nu_2}
}{|q_\bot|^5 |k_\bot|^3} +
\frac{g^\bot_{\mu_1 \nu_1} q^\bot_{\mu_2} k^\bot_{\nu_2}
+g^\bot_{\mu_2 \nu_1} q^\bot_{\mu_1} k^\bot_{\nu_2}
+g^\bot_{\mu_1 \nu_2} q^\bot_{\mu_2} k^\bot_{\nu_1}
+g^\bot_{\mu_2 \nu_2} q^\bot_{\mu_1} k^\bot_{\nu_1}
}{|q_\bot|^3 |k_\bot|^3}
\nonumber
\\ &&
+ \frac{q^\bot_{\mu_1} k^\bot_{\mu_2} g^\bot_{\nu_1 \nu_2}
+ k^\bot_{\mu_1} q^\bot_{\mu_2} g^\bot_{\nu_1 \nu_2}
}{|q_\bot|^3 |k_\bot|^3}
+ \frac{q^\bot_{\nu_1} k^\bot_{\nu_2} g^\bot_{\mu_1 \mu_2}
+k^\bot_{\nu_1} q^\bot_{\nu_2} g^\bot_{\mu_1 \mu_2}
}{|q_\bot|^3 |k_\bot|^3}
-3z \frac{q^\bot_{\mu_1} q^\bot_{\mu_2} g^\bot_{\nu_1 \nu_2}}{|q_\bot|^4 |k_\bot|^2}
-3z \frac{k^\bot_{\nu_1} k^\bot_{\nu_2} g^\bot_{\mu_1 \mu_2}}{|q_\bot|^2 |k_\bot|^4}
\nonumber
\\ &&
+ z \frac{g^\bot_{\mu_1 \mu_2} g^\bot_{\nu_1 \nu_2}}{|q_\bot|^2 |k_\bot|^2}\,.
\ee
This tensor is symmetric  both under
$\mu_1 \leftrightarrow \mu_2$, $\nu_1 \leftrightarrow \nu_2$
permutations
and $q_\bot \leftrightarrow k_\bot $
interchange and satisfies
\be
\frac{d^4z}{dq_{\bot}^{\mu_1 } dq_{\bot}^{\mu_2 } dk_{\bot}^{\nu_1 } dk_{\bot}^{\nu_2 }}
k_{\bot}^{\nu_2}=-
\frac{d^3z}{dq_{\bot}^{\mu_1 } dq_{\bot}^{\mu_2 } dk_{\bot}^{\nu_1 } }.
\ee

\end{document}